\documentclass[aps,prb,reprint,superscriptaddress,amsmath,amssymb,floatfix]{revtex4-2}
\usepackage{graphicx}
\usepackage{dcolumn}
\usepackage{bm}
\usepackage{xcolor}
\usepackage[colorlinks,linkcolor=blue,anchorcolor=blue,citecolor=blue,urlcolor=blue]{hyperref}
\definecolor{darkred}{rgb}{0.5, 0.0, 0.0}
\definecolor{darkgreen}{rgb}{0.0, 0.5, 0.0}

\usepackage{soul}

\begin{document}
\title{Scalable tensor network algorithm for thermal quantum many-body systems in two dimension}
\author{Meng Zhang}\thanks{The two authors made equal contributions.}
\affiliation{Institute of Artificial Intelligence, Hefei Comprehensive National Science Center, Hefei, 230088, People's Republic of China}

\author{Hao Zhang}\thanks{The two authors made equal contributions.}
\affiliation{CAS Key Laboratory of Quantum Information, University of Science and Technology of China, Hefei 230026, People's Republic of China}
\affiliation{Synergetic Innovation Center of Quantum Information and Quantum Physics, University of Science and Technology of China, Hefei 230026, China}

\author{Chao Wang}
\affiliation{Institute of Artificial Intelligence, Hefei Comprehensive National Science Center, Hefei, 230088, People's Republic of China}

\author{Lixin He}
\email{helx@ustc.edu.cn}
\affiliation{CAS Key Laboratory of Quantum Information, University of Science and Technology of China, Hefei 230026, People's Republic of China}
\affiliation{Institute of Artificial Intelligence, Hefei Comprehensive National Science Center, Hefei, 230088, People's Republic of China}
\affiliation{Synergetic Innovation Center of Quantum Information and Quantum Physics, University of Science and Technology of China, Hefei 230026, China}
\affiliation{Hefei National Laboratory, University of Science and Technology of China, Hefei 230088, China}

\date{\today}
\begin{abstract}
Simulating strongly-correlated quantum many-body systems at finite temperatures is a significant challenge in computational physics. In this work, we present a scalable finite-temperature tensor network algorithm for two-dimensional quantum many-body systems. We employ the (fermionic) projected entangled pair state (PEPS) to represent the vectorization of the quantum thermal state and utilize a stochastic reconfiguration method to cool down the quantum states from infinite temperature. We validate our method by benchmarking it against the 2D antiferromagnetic Heisenberg model, the $J_1$-$J_2$ model, and the Fermi–Hubbard model, comparing physical properties such as internal energy, specific heat, and magnetic susceptibility with results obtained from stochastic series expansion (SSE), exact diagonalization, and determinant quantum Monte Carlo (DQMC).
\end{abstract}
\maketitle

\section{Introduction}
Strongly correlated systems in two dimensions can host many intriguing phenomena, including quantum spin liquids \cite{Savary_2017,RevModPhys.89.025003,doi:10.1126/science.aay0668} and high-temperature superconductivity \cite{Bednorz1986,Keimer2015}, among others. Quantum Monte Carlo (QMC) methods have achieved great success in studying certain quantum  many-body systems in 2D, but often suffer from the sign problem when investigating frustrated and fermionic systems.
Developing novel numerical methods to explore strongly correlated systems in two dimensions is crucial for modern condensed matter physics.

 In recent years, tensor network methods have become powerful tools for dealing with frustrated systems and fermionic systems in two dimensions \cite{doi:10.1080/14789940801912366,ORUS2014117,RevModPhys.93.045003}.
 By capturing the entanglement structure of quantum many-body systems, tensor network methods enable the design of efficient and faithful representations of these systems. Tensor network methods have achieved fruitful results in simulating two-dimensional strongly correlated systems, which, however, have so far been applied primarily at zero temperature.

However, real physical systems always exist at finite temperatures. Furthermore, much richer physical phenomena emerge at finite temperatures. Therefore, it is necessary to develop tensor network methods that can simulate quantum many-body systems at finite temperatures.
Several works \cite{PhysRevB.92.035152,PhysRevLett.122.070502,PhysRevX.8.031082,PhysRevLett.130.226502,PhysRevLett.102.190601,Stoudenmire_2010,PhysRevX.11.031007,schmoll2022finite} have made progress in using tensor networks to handle thermal quantum many-body systems in two dimensions. One approach is based on infinite projected entangled pair states (iPEPS) or infinite projected entangled pair operators (iPEPO) to simulate thermal states in two dimensions \cite{PhysRevB.92.035152,PhysRevLett.122.070502,schmoll2022finite}.
This approach simulates thermal systems in the thermodynamic limit by imposing translational invariance on the PEPS. However, this constraint prevents it from simulating long-range interactions and systems with broken translational invariance.
Another approach uses one-dimensional matrix product states (MPS) or matrix product operators (MPO) to simulate quantum states in two dimensions \cite{PhysRevX.8.031082,PhysRevLett.130.226502,PhysRevLett.102.190601,Stoudenmire_2010,PhysRevX.11.031007}. However, to satisfy the area law in two dimensions, the parameters of MPS required to faithfully represent 2D quantum states increase exponentially as the width of the system increases, which limits this method to quasi-one-dimensional systems.

The combination of PEPS with the variational Monte Carlo (VMC) method, i.e. PEPS-VMC scheme \cite{PhysRevB.95.195154,PhysRevB.103.235155,PhysRevB.104.235141}
provides a scalable scheme to treat 2D strongly correlated systems. This approach has been used to solve the ground states for various quantum spin systems \cite{PhysRevB.98.241109,PhysRevX.12.031039,LIU20221034,PhysRevB.105.024411} and fermionic systems \cite{PhysRevB.99.195153,Dong2020}. Recent work \cite{lu2024variationalneuraltensornetwork} used various an{\"s}atze (including PEPS) combining with VMC to minimize modified free energy to obtain thermal state in  specific temperature, but the simulation of which only limited to Ising-type model.

In this work, we extend the PEPS-VMC scheme \cite{PhysRevB.95.195154,PhysRevB.103.235155,PhysRevB.104.235141} in a different approach to simulate thermal quantum states for both frustrated and Fermionic system in 2D. Utilizing vectorization \cite{NatCommunOrus,PhysRevLett.122.070502}, we can map the density matrix to a state vector, which can be represented by a PEPS.
By applying the imaginary-time evolution operator to the infinite temperature thermal state, a sequence of thermal states can be generated. Unlike prior work \cite{PhysRevLett.122.070502,schmoll2022finite}, which utilizes a simple update scheme for the imaginary-time evolution, we adopt the stochastic reconfiguration (SR) method \cite{PhysRevLett.80.4558,PhysRevB.64.024512}. The SR method, which is equivalent to imaginary-time projection \cite{Becca_Sorella_2017,doi:10.7566/JPSJ.85.034601,PhysRevLett.127.060601,PhysRevB.106.165111,nys2024realtimequantumdynamicsthermal}, allows for a variational update of the PEPS, leading to more accurate thermal state evolution.
This approach also inherits the advantages of the PEPS-VMC scheme used in ground state calculations \cite{PhysRevB.95.195154,PhysRevB.103.235155,PhysRevB.104.235141}, which can be efficiently implemented using massive parallelization. Additionally, it only requires handling single-layer tensor networks for both optimization and the calculation of observables, significantly simplifying the computational process.

 We benchmark our method on quantum spin models and the Hubbard model by comparing it with other numerical techniques, including stochastic series expansion (SSE) \cite{PhysRevB.43.5950,AWSandvik_1992}, exact diagonalization (ED), and determinant quantum Monte Carlo (DQMC) \cite{PhysRevD.24.2278,PhysRevB.28.4059,PhysRevB.31.4403}. The results obtained using our approach show excellent agreement with these established methods, demonstrating its accuracy and reliability.

\section{Method}\label{method}

Given a Hamiltonian $H$ of a quantum system, the (unnormalized) thermal state at temperature $T = \frac{1}{\beta}$ is expressed as $\rho_{\beta} = e^{-\beta H}$. This can be rewritten as:
\begin{equation}
\rho_\beta = e^{-\frac{\beta}{2} H} I e^{-\frac{\beta}{2} H},
\end{equation}
where $I$ is the identity operator, corresponding to the infinite temperature thermal state (i.e., $\beta = 0$).

For each density matrix $\rho$, we can apply the following vectorization transformation to map it into a vector $\left| \rho \right\rangle_{\sharp}$ in the Hilbert space $V \otimes V$:
\begin{equation}
\rho = \sum_{ss'} \rho_{ss'} |s\rangle \langle s'| \rightarrow \left| \rho \right\rangle_{\sharp} = \sum_{ss'} \rho_{ss'} |s\rangle |s'\rangle.
\end{equation}
Applying this transformation to the thermal state $\rho_{\beta}$, we obtain its vectorized form:
\begin{equation}
|\rho_\beta\rangle_{\sharp} = e^{-\frac{\beta}{2} H \otimes I} e^{-\frac{\beta}{2} I \otimes H^{T}} |I\rangle_{\sharp} = e^{-\frac{\beta}{2} \mathcal{H}} |I\rangle_{\sharp}.
\end{equation}
Here, $|I\rangle_{\sharp}$ represents the vectorized infinite temperature thermal state, and $\mathcal{H} = H \otimes I + I \otimes H^{T}$. Therefore, the thermal state at temperature $T = 1/\beta$ can be obtained by performing imaginary-time evolution starting from the infinite temperature thermal state.

\begin{figure}[htb]
\includegraphics[width=0.48\textwidth]{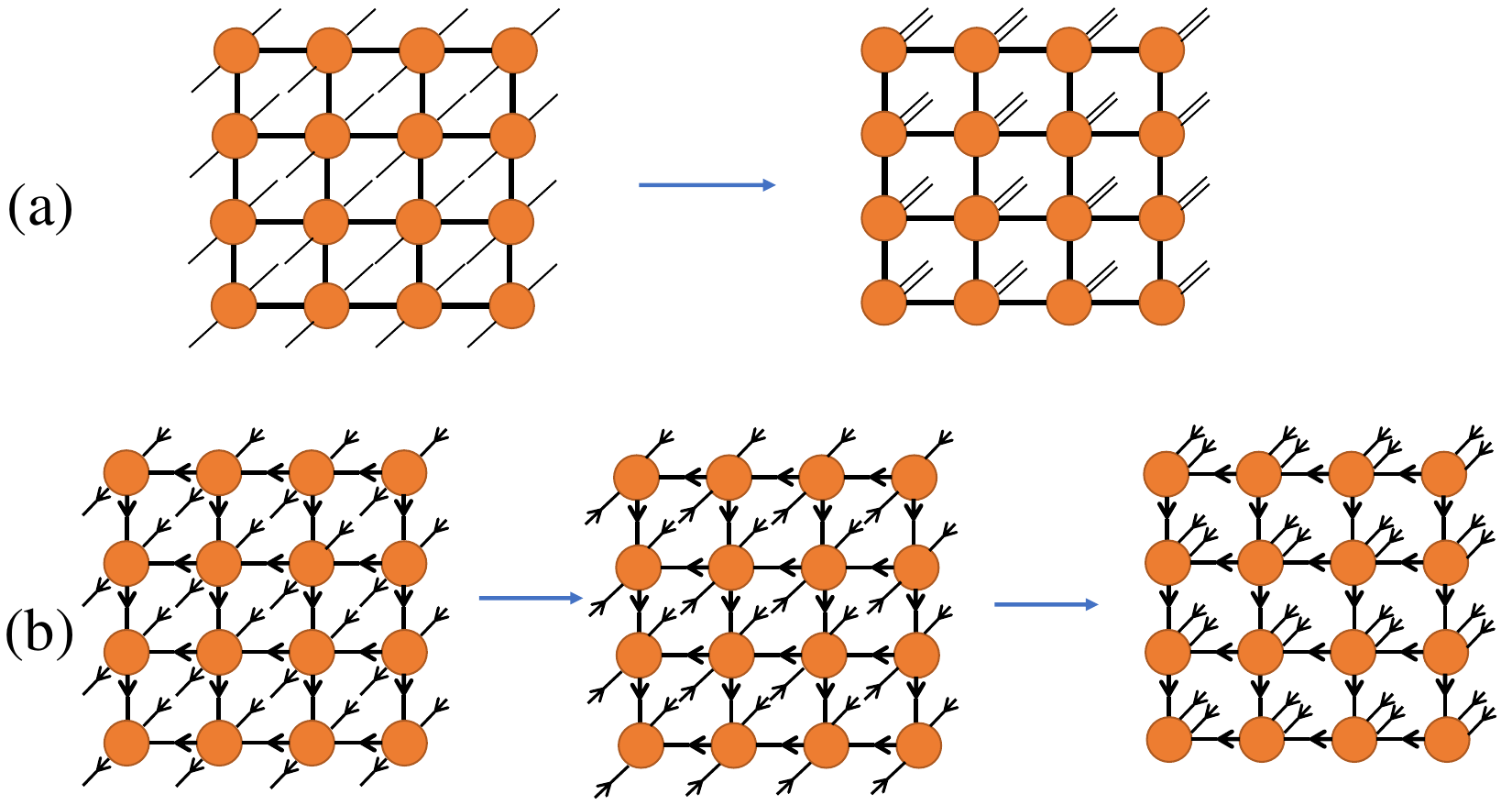}
\caption{ (a) Vectorization of PEPS: The two physical indices of each tensor are  treated as a single physical index with dimension $d^2$.
(b) Vectorization of fPEPS: First, the Fermi arrows of one of the physical indices for each fermionic tensor are reversed, after which the two physical indices are treated as a single physical index.
}
\label{tns}
\end{figure}

We use a PEPO to represent the density matrix. The vectorization of the PEPO is illustrated in Fig.~\ref{tns}(a), where the two physical indices can be regarded as a single physical index with dimension $d^2$. After vectorization, the density matrix can be expressed as a PEPS:
\begin{equation}
\left|\rho\right\rangle_{\sharp} = \sum_{S_{1}, \cdots, S_{N}=0}^{d^2-1} \operatorname{Tr}\left(T_{1}^{S_{1}} \cdots T_{N}^{S_{N}}\right) \left|S_{1} \cdots S_{N}\right\rangle,
\end{equation}
where $|S_{i}\rangle = |s_i\rangle |s'_{i} \rangle$, which has a dimension of $d^2$ in the local Hilbert space. $T_{i}^{S_{i}}$ is the tensor at the $i$-th site of the square lattice with $N = L_{1} \times L_{2}$ sites, where $N$ is the total number of sites.

The operator $e^{-\frac{\beta}{2}\mathcal{H}}$ can be further decomposed into smaller imaginary-time evolution operators, i.e., $e^{-\frac{\beta}{2}\mathcal{H}} = \prod_{i} e^{-\tau_{i} \mathcal{H}}$. To implement the imaginary-time evolution, conventional approaches typically utilize either simple update \cite{PhysRevLett.101.090603} or full update schemes \cite{PhysRevLett.101.250602} to update the tensor network. These methods require $e^{-\tau \mathcal{H}}$ to be approximately decomposed into smaller operators via the Suzuki-Trotter decomposition.

In this work, instead of using the aforementioned update schemes, we employ the SR method \cite{PhysRevLett.80.4558, PhysRevB.64.024512} to variationally update the parameters of the tensor network. We assume that the variational (unnormalized) vectorized density matrix, $\left| \rho (\theta)\right\rangle_{\sharp} = \sum_{S} \rho_{\theta}(S)\left| S \right\rangle$, is represented as a PEPS with parameters $\theta = (\theta_{1}, \theta_{2}, \dots, \theta_{N_p})$, where $N_p$ is the number of
parameters.
Starting from the initial state $|I\rangle_{\sharp}$, at each step of the SR method, we performed an update to the vectorized density matrix according to $\left | \rho _{i+1} (\theta^{\prime}) \right \rangle_{\sharp } \simeq e^{-\tau\mathcal{H}}\left | \rho _{i} (\theta)\right \rangle_{\sharp }$. This was achieved by adjusting the variational parameters so that the updated state $\left | \rho_{i+1} (\theta^{\prime}) \right \rangle_{\sharp }$ closely approximates $e^{-\tau\mathcal{H}}\left | \rho _{i} (\theta)\right \rangle_{\sharp }$, for which the parameters are updated according to the following equation:
\begin{equation}
    \theta^{\prime} = \theta - \tau G^{-1}g,
\end{equation}
where the metric $G$ is given by
\begin{equation}\label{metric}
    G_{kk^{\prime}} =  \langle \overline{O_{k}(S)}O_{k^{\prime }}(S) \rangle - \langle \overline{O_{k}(S)}\rangle \langle O_{k^{\prime }}(S)  \rangle,
\end{equation}
and the gradient $g$ is
\begin{equation}\label{gradient}
    g_{k}= \langle \overline{O_{k}(S)} E_{\text{loc}}(S) \rangle - \langle \overline{O_{k}(S)}\rangle \langle E_{\text{loc}}(S) \rangle.
\end{equation}
Here, $\left \langle \dots \right \rangle$ denotes the average over configurations $S$ with weight $|\rho_{\theta}(S)|^{2}/_{\sharp}\left \langle \rho(\theta) | \rho(\theta) \right \rangle_{\sharp }$, $\overline{\dots}$ indicates complex conjugation, and the variables $O_{k}(S)$ and $E_{\text{loc}}(S)$ are defined as
\begin{eqnarray}
O_{k}(S) & = &\frac{1}{\rho_{\theta }(S)} \frac{\partial \rho_{\theta }(S) }{\partial \theta _{k}},\\
E_{\text{loc}}(S) & =  & \sum_{S^{\prime}}\frac{\rho_{\theta }(S^{\prime})}{\rho_{\theta }(S)} \mathcal{H}_{SS^{\prime }} .
\end{eqnarray}
Equations (\ref{metric}) and (\ref{gradient}) can be calculated by Monte Carlo sampling of the state $\left | \rho(\theta) \right \rangle_{\sharp }$.

To numerically obtain the vector $\delta \theta = G^{-1}g$, where $G$ is a Hermitian matrix, we employ the conjugate gradient (CG) method. The CG method is advantageous because it only requires the action of the matrix on a vector rather than the explicit construction of the entire $G$ matrix. This allows for efficient handling of large systems where constructing $G$ would be computationally expensive. We first rewrite the expression for the matrix elements $G_{kk'}$ as follows:
\begin{equation}\label{matrix-metric}
G_{kk^{\prime}} =  \frac{1}{N_s} \sum_{i}\overline{O_{ik}}O_{ik^{\prime }} - \frac{1}{N_s} \sum_i \overline{O_{ik}}\frac{1}{N_s} \sum_i O_{ik^{\prime }},
\end{equation}
where $i$ is the index of sampled configurations.

We do not  directly calculate $\delta \theta_{k} G_{kk'} \delta \theta_{k'}$. Instead, we only compute $O_{i k'} \delta \theta_{k'}$ and $\delta \theta_{k} \overline{O_{i k}}$ individually, which is a common technique used in the CG method for solving normal equations \cite{Bjrck1979,Paige1982}.
Applying the matrix to a vector incurs a computational cost of $O(N_p N_s)$, where $N_p$ is the number of parameters
to be optimized and $N_s$ is the number of samples. Consequently, the total time complexity is $O(N_p N_s N_{cg})$, where $N_{cg}$ is the number of CG iterations. Typically, a few hundred iterations are sufficient for convergence.
In our case, $N_s \ll N_p$, with $N_s \approx 10^4$ and $N_p \approx 10^6$, therefore applying $G_{kk'}$ using
this separable form is much more efficient than using the full matrix form, which has a time complexity of $O(N_p^2 N_{cg})$.

The internal energy can be calculated as,
\begin{equation}\label{in_en}
    E(2\beta) = \frac{_{\sharp}\left\langle\rho_\beta \right| \mathcal{H}\left | \rho_\beta \right \rangle_{\sharp }}
    {2_{\sharp}\left \langle \rho_\beta  | \rho_\beta  \right \rangle_{\sharp } }
    = \frac{\text{Tr}\left (\rho_\beta \rho_\beta^{\dagger}H+ \rho_\beta^{\dagger}\rho_\beta H \right )}{2\text{Tr}\left (\rho_\beta \rho_\beta^{\dagger}  \right ) } ,
\end{equation}
which is the internal energy  of sates $\left (\rho_\beta \rho_\beta^{\dagger}+\rho_\beta^{\dagger}\rho_\beta  \right )/\left(2\text{Tr}\left (\rho_\beta \rho_\beta^{\dagger}  \right ) \right)$ that can be regarded as the thermal state with inverse temperature $2\beta$.
Equation~(\ref{in_en}) can be calculated using the Monte Carlo average of $E_{\text{loc}}(S)$. Thus, we only need to contract a single layer tensor network $\rho_{\theta } (S)$.  Other thermal observations such as susceptibility can be calculated in similar way.

To sample the quantum state $\left | \rho_\beta \right \rangle_{\sharp }$ represented by PEPS, we can employ the Markov Chain Monte Carlo (MCMC) method \cite{PhysRevB.103.235155} or the direct sampling method \cite{PhysRevB.104.235141} for PEPS. However, at very high temperatures, the (unnormalized) thermal state $\rho_\beta$ is close to the identity operator, where the diagonal entries are close to one and the off-diagonal entries are close to zero. As a result, it becomes difficult to obtain the important configurations $|s\rangle|s^{\prime}\rangle$ ($s \ne s^{\prime}$) through sampling at high temperatures, which is important for calculating the energy and gradients.
To overcome this, we employ a reweighting sampling method. Instead of sampling the quantum state $\left | \rho _{\beta} \right \rangle_{\sharp }$ according to the weight $|\rho_{ss^{\prime}}|^{2}$, we sample according to its square root, i.e., $|\rho_{ss^{\prime}}|$. This approach allows more configurations $|s\rangle|s^{\prime}\rangle$ ($s \ne s^{\prime}$) to be sampled by the Markov chain. Alternatively, one can evolve the quantum states at high temperatures using the SU method \cite{PhysRevLett.101.090603}. For more details, see Appendix \ref{sec:hightemp}.

The finite temperature method described above can also be applied to fermionic systems through the use of fermionic PEPS (fPEPS) \cite{PhysRevB.81.165104,PhysRevB.84.041108,Gu2010,PhysRevB.88.155112,PhysRevA.81.052338,PhysRevB.99.195153}.
The ``Fermi arrows'' were introduced for fPEPS to specify the order of the fermion operators in the entangled EPR pairs\cite{PhysRevB.99.195153}.
The vectorization of the density matrix for a fermionic system is illustrated in Fig.~\ref{tns}(b). First, we reverse the Fermi arrows of one of the physical indices  in the conjugate space so that the two physical indices can be treated as a single physical index. Additionally, the arrows on the corresponding edges in the operator $H^T$ are also reversed according to the operation rules\cite{PhysRevB.99.195153,ZHANG2024109355}.

\section{Benchmark Results}
\subsection{Quantum Spin Models}

We first benchmark our method with 2D antiferromagnetic Heisenberg model and 2D $J_{1}-J_{2}$ model on a square lattice, which can be written as,
\begin{equation}\label{j1j2eq}
    H = J_{1} \sum_{\langle i, j\rangle} \mathbf{S}_{i} \cdot \mathbf{S}_{j} + J_{2} \sum_{\langle\langle i, j\rangle\rangle} \mathbf{S}_{i} \cdot \mathbf{S}_{j},
\end{equation}
where $\mathbf{S}_{i}$ is the spin-1/2 operator at site $i$ and $\langle i, j\rangle$ represents the nearest neighbor pairs  and $\langle\langle i, j\rangle\rangle$ represents the next-nearest-neighbor pairs.

\begin{figure}[htb]
\includegraphics[width=0.48\textwidth]{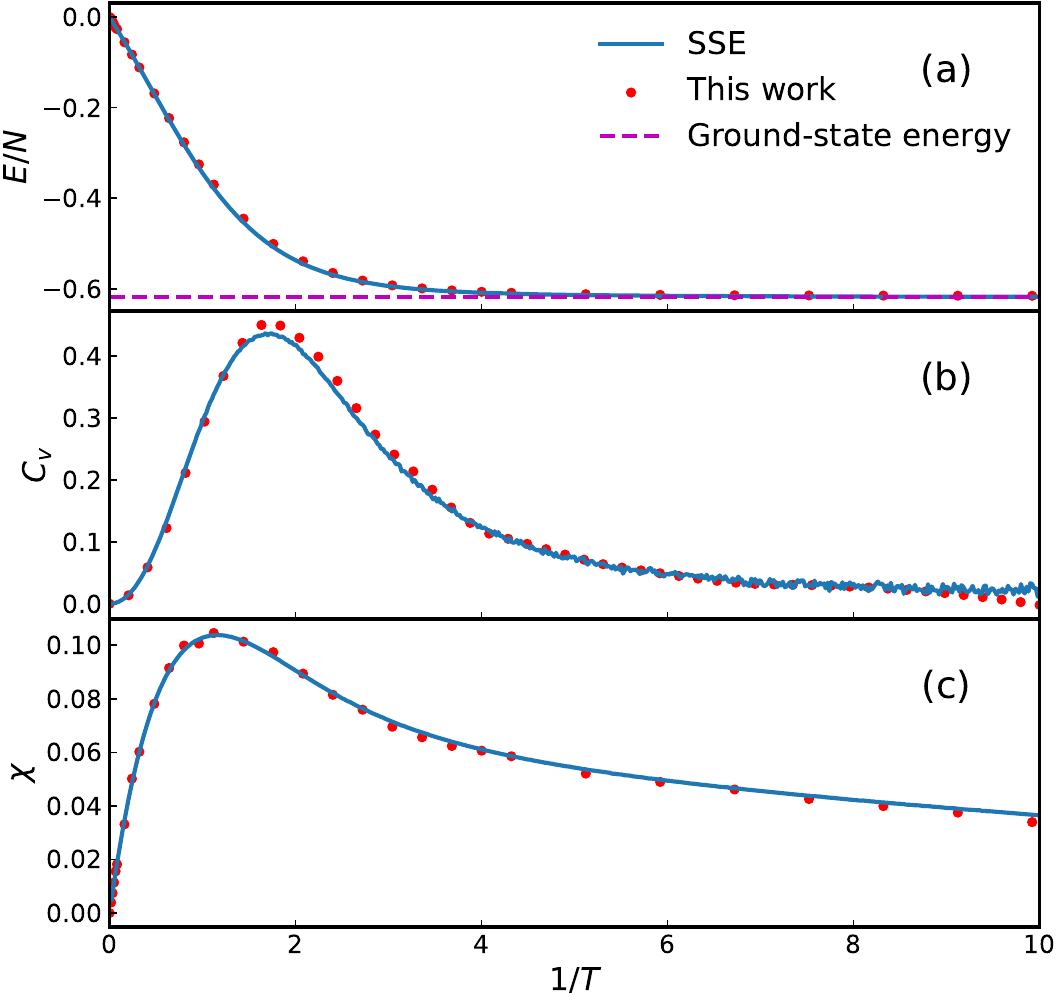}
\caption{\label{Heisenberg} Comparison of PEPS results (red dots) with SSE results (blue line) for a 2D Heisenberg model on an $8 \times 8$ square lattice: (a) Energy per site, (b) Specific heat per site, and (c) Uniform magnetic susceptibility $\chi$ as functions of $\beta = 1/T$. The dashed line in (a) indicates the ground state energy obtained by PEPS with $D=10$ of Ref.\cite{PhysRevB.95.195154}.}
\end{figure}

When $J_{1}=1$ and $J_{2}=0$, Eq.~(\ref{j1j2eq}) reduces to the non-frustrated Heisenberg model, which can be simulated using the SSE method without suffering from the sign problem \cite{PhysRevB.43.5950,AWSandvik_1992}.
As shown in Fig.~\ref{Heisenberg}, we compare the results obtained by our method with those from the SSE method. The size of the square lattice is $N = 8 \times 8$ with open boundary conditions (OBC). We set the bond dimension of the PEPO to $D=8$ and use the boundary-MPS contraction method \cite{doi:10.1080/14789940801912366,PhysRevB.103.235155} to contract the single-layer tensor network $\rho_{\theta}(S)$. The accuracy of the contraction is controlled by the boundary dimension $D_c$, which we set to $D_c = 16$. We adopt MCMC to sample the quantum state, with a total of 4$\times 10^4$ samples.
For high temperatures, we use the reweighting sampling method mentioned in Sec.~\ref{method} until $\beta \simeq 0.3$, and then we switch to normal sampling for lower temperatures down to $\beta \simeq 10$. We modified the boundary conditions of the SSE code \cite{SSEcode} to OBC to compare with our results. For the SSE method, we used $10^7$ samples.

As shown in Fig.~\ref{Heisenberg}(a), the internal energy per site calculated by our method agrees with the results of the SSE method across the entire temperature range $\beta = [0, 10]$, indicating that our method possesses excellent precision even at very low temperatures.

Figure~\ref{Heisenberg}(b) shows the specific heat $C_v$ as a function of $\beta$. To obtain $C_v$, we fit the energy data using a B-spline function and then calculate $C_v$ through numerical differentiation of the fitted curve.
As shown in Fig.~\ref{Heisenberg}(b), our results are in very good agreement with those of the SSE method.
Although the peak value of $C_v$ does not perfectly match the SSE results, primarily due to the errors inherent in numerical differentiation, the corresponding temperature agrees very well with the result of SSE.

Figure~\ref{Heisenberg}(c) shows how the magnetic susceptibility changes with $\beta$, which is defined as
\begin{equation}\label{sus}
    \chi = \frac{\beta}{N}\left( \left\langle M_{z}^{2}\right\rangle - \left\langle M_{z}\right\rangle^{2}\right), \quad M_{z} = \sum_{i=1}^{N} S_{i}^{z}.
\end{equation}
We set $\left\langle M_{z}\right\rangle^{2} = 0$ following the method in \cite{10.1063/1.3518900}. As shown in Fig.~\ref{Heisenberg}(c), the uniform magnetic susceptibility agrees well with the SSE results across the entire temperature range. Calculating susceptibility is more challenging than magnetization at low temperatures because, as the temperature decreases, the $\beta$ in Eq.~(\ref{sus}) increases, requiring higher accuracy in estimating $\left\langle M_{z}^{2}\right\rangle$ to achieve a fixed precision for susceptibility. This requirement can account for the slight deviation in susceptibility from the SSE results in the low-temperature region around $\beta \simeq 10$.

\begin{figure}[htb]
\includegraphics[width=0.48\textwidth]{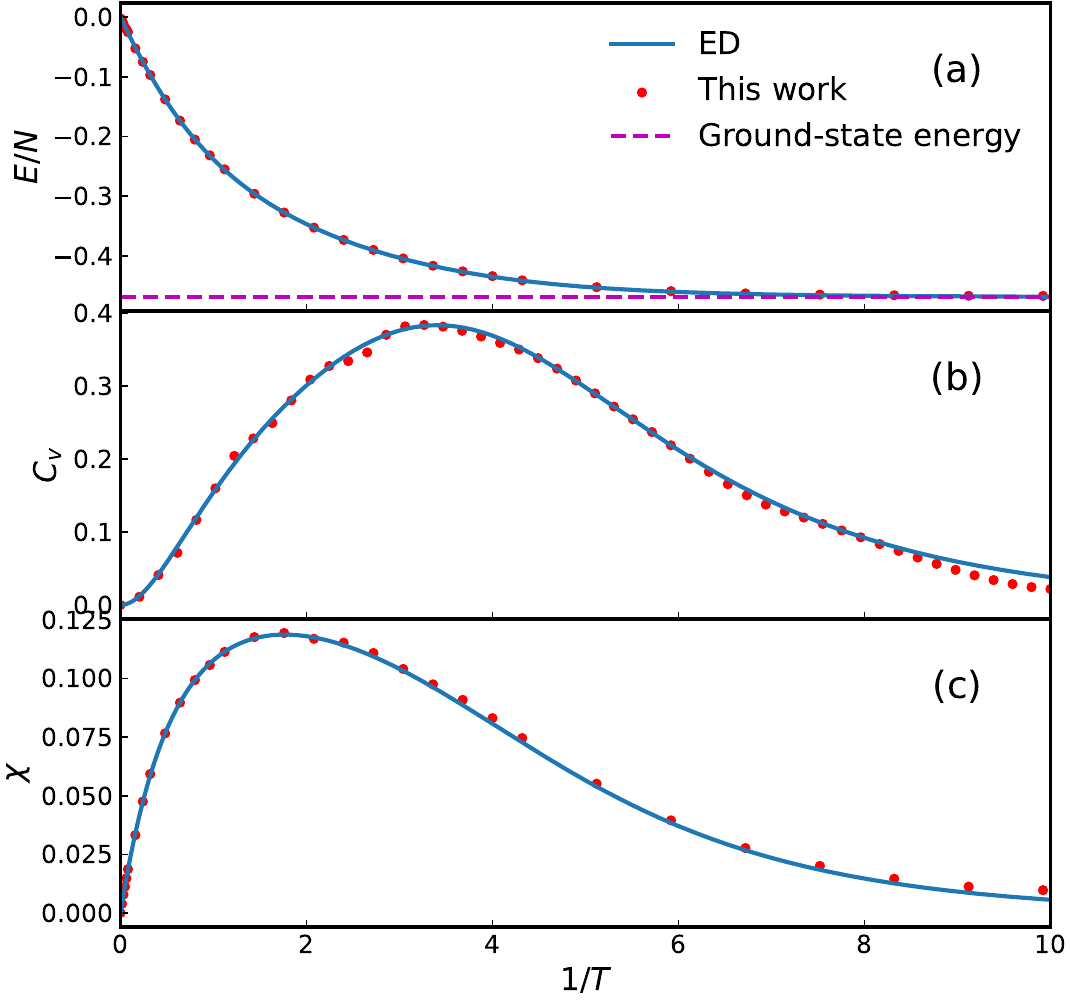}
\caption{\label{j1j2} Comparison of PEPS results (red dots) with ED results (blue line) for a 2D $J_{1}$-$J_{2}$ model on a $4 \times 4$ square lattice: (a) Energy per site, (b) Specific heat per site, and (c) Uniform magnetic susceptibility $\chi$ as functions of $\beta = 1/T$. The dashed line in (a) indicates the ground state energy obtained by ED.}
\end{figure}

When $J_{2} \neq 0$ in Eq.~(\ref{j1j2eq}), the $J_{1}-J_{2}$ model becomes a frustrated model, making it unsuitable for calculation using the QMC method due to the sign problem.
At zero temperature, many numerical studies \cite{PhysRevB.86.024424,PhysRevB.88.060402,PhysRevLett.121.107202,PhysRevB.98.241109,PhysRevX.12.031039,LIU20221034} have shown that the $J_{1}-J_{2}$ model hosts a quantum spin liquid phase when $J_{2}/J_{1} = 0.5$. We calculated the $J_{1}-J_{2}$ model by setting $J_{2}/J_{1} = 0.5$ on a $4 \times 4$ square lattice, which can be exactly solved using the ED method. We set the bond dimension of the PEPO to $D = 8$ and the boundary dimension to $D_c = 16$, with the number of samples set to 8$\times 10^4$.

As demonstrated in Fig.~\ref{j1j2}(a), the internal energy per site exhibits excellent agreement with the ED results across the entire temperature range. Similarly, the specific heat $C_v$, as shown in Fig.~\ref{j1j2}(b), closely matches the ED data. The magnetic susceptibility, depicted in Fig.~\ref{j1j2}(c), also aligns well with the ED results. These findings underscore the capability of our method to accurately simulate highly frustrated systems at finite temperatures.

\subsection{Fermi Hubbard model}

We benchmark our method by studying the Fermi Hubbard model on a square lattice at finite temperature,
\begin{equation}\label{equ:hubbard_model}
    H = -t \sum_{\langle i, j\rangle, \sigma} c_{i\sigma}^\dag c_{j\sigma} + U \sum_{i} n_{i\uparrow} n_{i\downarrow} - \mu \sum_{i\sigma} n_{i\sigma},
\end{equation}
where $c_{i\sigma}$ is the annihilation operator at site $i$ with spin $\sigma = \uparrow$ or $\downarrow$, and $n_{i\sigma}=c_{i\sigma}^\dag c_{i\sigma}$ is the particle number operator.
$U$ is the on-site Coulomb repulsion, and $\mu$ is the chemical potential.

In the calculations, we use a $6 \times 6$ square lattice with OBC. The parameters of the model are set to $t=1$, $U=8$, and $\mu=4$, corresponding to a half-filled antiferromagnetic phase at ground state\cite{Zheng2016}. While our method is also applicable to the canonical ensemble, we use the grand canonical ensemble here for comparison with DQMC. Since this is the half-filling phase, there is no sign problem for DQMC.

\begin{figure}
    \centering
    \includegraphics[width=0.42\textwidth]{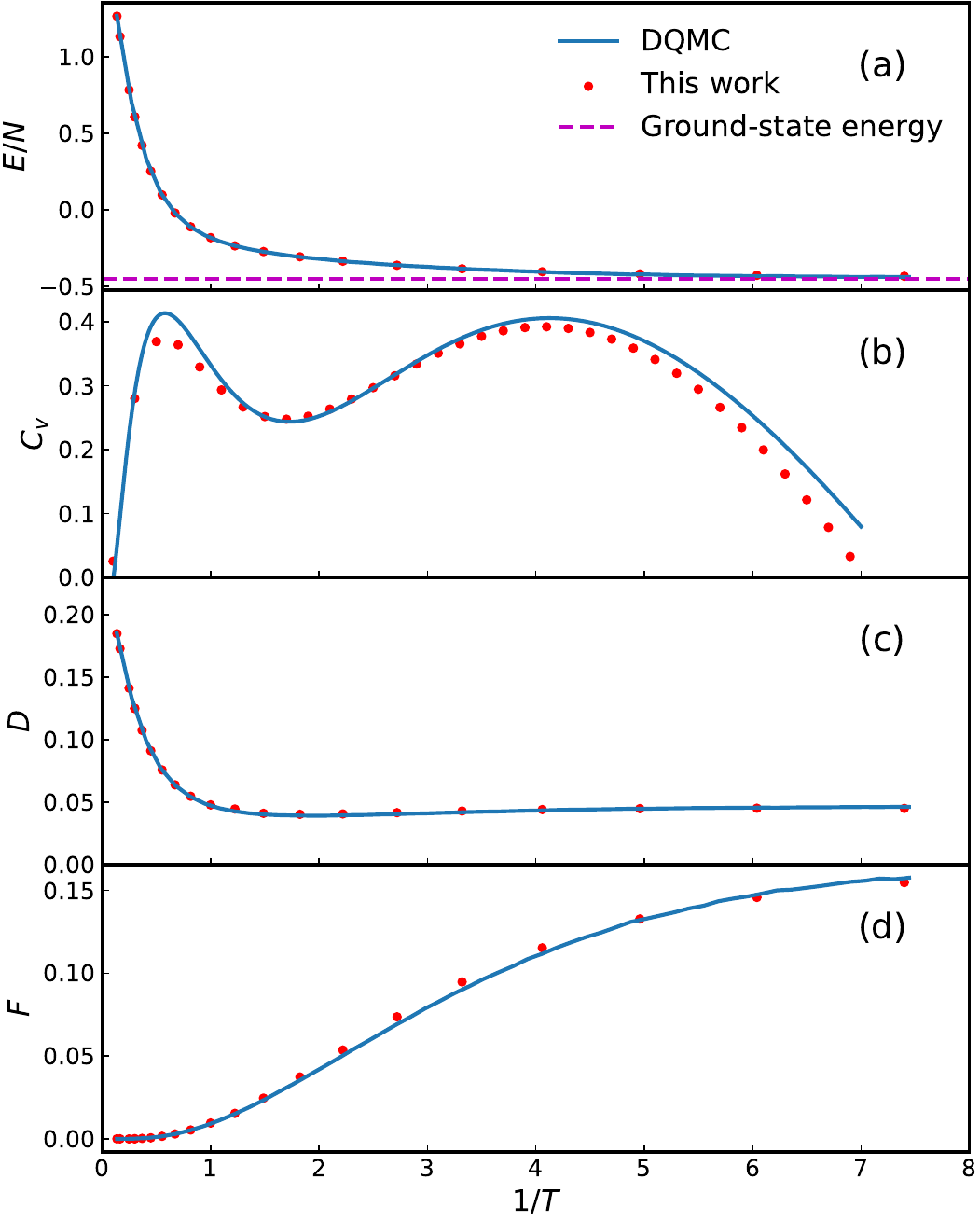}
    \caption{Comparison of PEPS results (red dots) with DQMC results (blue line) for a Fermi-Hubbard model on a $6 \times 6$ square lattice: (a) Energy per site, (b) Specific heat per site, (c) Double occupancy per site, and (d) Spin correlation between next-nearest neighbor sites, all as functions of $\beta = 1/T$. The dashed line in (a) indicates the ground state energy which obtained by by PEPS with $D=16$. The model parameters are set to $U=8$, $t=1$, and $\mu=4$.}
    \label{fig:6x6-hubbard}
\end{figure}

We compare our results for the Hubbard model with those obtained using the DQMC method in Fig. \ref{fig:6x6-hubbard}.
We use a modified direct sampling method \cite{PhysRevB.104.235141}, detailed in Appendix \ref{sec:hightemp}, with up to 3$\times 10^4$ samples during optimization and 6$\times 10^5$ samples during measurement.

In Fig.~\ref{fig:6x6-hubbard}(a), we show the internal energy per site $E/N$ as it decreases with temperature, approaching the ground state energy indicated by the dashed line. Our results are in excellent agreement with those obtained using the DQMC method.

The heat capacity $C_v$ is calculated using Gaussian process regression combined with numerical differentiation, as shown in Fig.~\ref{fig:6x6-hubbard}(b). While there is a small discrepancy compared to the DQMC results, this difference likely arises due to the higher computational precision required for accurately calculating $C_v$, which may necessitate a larger bond dimension $D$.
Despite this, two distinct peaks in $C_v$ are observed at approximately $1/T_1 = 0.5$ and $1/T_2 = 4.0$, which are consistent with the DQMC results and previously reported findings in Ref.~\cite{PhysRevLett.130.226502}. The high-temperature peak is attributed to the excitation of the Hubbard bands, while the low-temperature peak is associated with the excitation of spin states \cite{PhysRevB.55.12918}. The high-temperature peak has been estimated as $1/T_1 \approx \frac{4.8}{U} = 0.6$, and the low-temperature peak as $1/T_2 \approx \frac{3U}{8t^2} = 3.0$ \cite{PhysRevB.63.125116}. Our results are in good agreement with these estimates.

We also calculate the double occupancy per site, $D = \sum_i \langle n_{i\downarrow} n_{i\uparrow} \rangle / N$, and the spin correlation between next-nearest neighbor sites, $F = \sum_{\langle\langle i,j\rangle\rangle}\langle \mathbf{S}_i \cdot \mathbf{S}_j \rangle / M$, where $M$ is the number of next-nearest neighbor pairs. The results are shown in Fig.~\ref{fig:6x6-hubbard}(c) and Fig.~\ref{fig:6x6-hubbard}(d), respectively, and are in very good agreement with those obtained using the DQMC method.
The double occupancy is around 0.2 at very high temperatures and decreases to approximately 0.05 as the temperature approaches zero. Conversely, the spin correlation between next-nearest neighbor sites increases from zero at very high temperatures to about 0.15 at low temperatures.
Due to the small system size, finite-size effects prevent the observation of distinct discontinuities. Nonetheless, our results are in close agreement with those reported in Ref.~\cite{PhysRevLett.130.226502}, except for minor differences in the cooling curves, which may be due to the differences in the system size used in the simulations.

\section{Summary}

We develop a finite-temperature PEPS-VMC scheme to simulate thermal quantum many-body systems in 2D. The (f)PEPS representation used in this approach is scalable in two dimensions. We benchmark our method on quantum spin models and the Fermi-Hubbard model, and the results show very good agreement with exact solutions. Our approach can also be directly applied to simulate systems with long-range interactions, such as Rydberg atoms. The recently proposed superposed PEPS ansatz \cite{dong2024boundaries} can be used within our framework to simulate systems with periodic boundary conditions. This method provides a powerful tool for simulating strongly correlated quantum many-body systems at finite temperatures.

\begin{acknowledgments}
We thank Mingzhong Lu for providing the DQMC data, and Shaojun Dong, Yufeng Song, and Youjin Deng for valuable discussions. This work was supported by the National Natural Science Foundation of China under Grant Nos. 12134012 and 12304552, and by the Strategic Priority Research Program of the Chinese Academy of Sciences under Grant No. XDB0500201, as well as the Innovation Program for Quantum Science and Technology under Grant No. 2021ZD0301200. The numerical calculations in this study were performed on the ORISE Supercomputer and the USTC HPC facilities.

\end{acknowledgments}

\appendix
\section{Evolving Quantum States at very High Temperatures}
\label{sec:hightemp}

At infinite temperature, the thermal state is represented by the identity operator $I$, which can result in a vanishing gradient problem. To avoid this issue in the initial stages, we introduce small perturbations to the zero entries of the PEPS at infinite temperature, similar to the method used for neural networks in Ref.~\cite{PhysRevLett.127.060601}.
However, the off-diagonal terms in the density matrices remain very small, preventing some important spin configurations from being sampled, even though they are crucial for accurately computing the gradients. To address this challenge, we adopt a reweighting sampling method at very high temperatures.

In the standard MCMC approach, the (unnormalized) sampling probability is  $p_0(S) = |\rho(S)|^2$, where $S$ represents the spin configuration (for simplicity, we denote $ss'$ as  $S$). However, at very high temperature, the off-diagonal elements of the density matrix are very small. Squaring these small off-diagonal elements results in even smaller values, making the corresponding spin configurations difficult to sample effectively. Therefore, instead of sampling according to $p_0(S)$, we perform sampling based on a modified probability distribution $p_1(S) = |\rho(S)|$. 
For quantity $f(S)$ that to be estimated, we have
\begin{equation}
     \left \langle f(S) \right \rangle  _{p_0} = \frac{\left \langle f(S) \cdot r_1(S) \right \rangle_{p_1} }{\left \langle r_1(S) \right \rangle_{p_1}},
\end{equation}
where the $\left \langle \dots  \right \rangle_{p_1}$ is the Monte Carlo average corresponding to probability $p_1(S)$, and $r_1(S) = p_0(S)/p_1(S) = |\rho(S)|$, and this distribution is achieved by Metropolis sampling.
We also notice that similar reweighting method have been adapted in Ref.\cite{chen2023efficient} in the task of optimizing the ground state wave function.

As for a Fermion system, it is even harder to sample, so we deploy the direct sampling method on it \cite{PhysRevB.104.235141}, where the configurations are sampled autoregressively according to:
\begin{equation}
p_2(S_1 S_2 S_3 \cdots) = p_{2c}(S_1) p_{2c}(S_2 \mid S_1) p_{2c}(S_3 \mid S_1 S_2) \cdots ,
\end{equation}
where
\begin{equation}
\begin{aligned}
&p_{2c}(S_i \mid S_1 S_2 \cdots S_{i-1}) \\
= &\frac{\sum_{S_{i+1} S_{i+2} \cdots} |\rho(S_1 S_2 \cdots S_{i-1} S_i S_{i+1} S_{i+2} \cdots)|^2}{\sum_{S_i S_{i+1} S_{i+2} \cdots} |\rho(S_1 S_2 \cdots S_{i-1} S_i S_{i+1} S_{i+2} \cdots)|^2}.
\end{aligned}
\end{equation}
Here, $p_{2c}(S_i \mid S_1 S_2 \cdots S_{i-1})$ is the conditional probability to sample the spin configuration at site $i$. As a result, we have $p_2(S) \propto p_0(S)$.
To apply the reweighting technique on direct sampling,
we first use a similar reweighting strategy to the conditional probability, i.e.,
\begin{equation}
p_{3c}(S_i \mid S_1 S_2 \cdots S_{i-1}) = \sqrt{p_{2c}(S_i \mid S_1 S_2 \cdots S_{i-1})}.
\end{equation}
In addition, to sample more crucial configurations, we introduce another reweighting constant over the conditional probability,
\begin{equation}
p_{4c}(S_i \mid S_1 S_2 \cdots S_{i-1}) = b(S_i) p_{3c}(S_i \mid S_1 S_2 \cdots S_{i-1}),
\end{equation}
where $b(s_i s'_i) = 1$ if $s_i \neq s'_i$, and otherwise, it is a dynamic value, which would be updated every gradient step to ensure that the number of sampled diagonal terms does not exceed half of the total number of samples.
Ultimately, we use the following probability distribution for sampling:
\begin{equation}
p_4(S_1 S_2 S_3 \cdots) = p_{4c}(S_1) p_{4c}(S_2 \mid S_1) p_{4c}(S_3 \mid S_1 S_2) \cdots ,
\end{equation}
and the observables are calculated according to:
\begin{equation}
\left \langle f(S) \right \rangle _{p_0} = \frac{\left \langle f(S) r_4(S) \right \rangle_{p_4}}{\left \langle r_4(S) \right \rangle_{p_4}},
\end{equation}
where $r_4(S) = \frac{p_0(S)}{p_4(S)}$.

An alternative approach to solving the gradient vanishing and sampling problem at high temperatures is to apply the SU method for evolution at these temperatures. At very high temperatures, the thermal state closely approximates a direct product state $|I\rangle_{\sharp}$, indicating that environmental effects are minimal. This makes the SU scheme~\cite{PhysRevLett.101.090603} particularly suitable for updating the tensor network at high temperatures. After several tens or hundreds of simple update steps, we transition to the SR update method, which provides greater accuracy at lower temperatures.

\nocite{*}
%
\end{document}